# Understanding Online Behaviors through a Temporal Lens

Tai-Quan Peng[1] (0000-0002-2588-7491), Jonathan J. H. Zhu[2,3] (0000-0001-6173-6941)

1. Department of Communication, Michigan State University, East Lansing, USA
2. Department of Media and Communication, City University of Hong Kong, Hong Kong
3. School of Data Science, City University of Hong Kong, Hong Kong

**Abstract:**
Timestamps in digital traces include significant detailed information on *when* human behaviors occur, which is universally available and standardized in all types of digital traces. Nevertheless, the concept of time is under-explicated in empirical studies of online behaviors. This paper discusses the (un)desirable properties of timestamps in digital traces and summarizes how timestamps in digital traces have been utilized in existing studies of online behaviors. The paper argues that time-in-behaviors perspective can provide a microscope with a renovated temporal lens to observe and understand online behaviors. Going beyond the traditional behaviors-in-time perspective, time-in-behaviors perspective enables empirical examination of online behaviors from multiple units of analysis (e.g., discrete behaviors, behavioral sessions, and behavioral trajectories) and from multiple dimensions (e.g., duration, order, transition, rhythm). The paper shows the potentials of the time-in-behaviors perspective with several empirical cases and proposes future directions in explicating the concept of time in computational social science.

# 1 Introduction

The volume, velocity, and variety of digital traces generated by millions of users have triggered the interest and creativity of social scientists to address old questions in new ways and to ask new questions that were out of reach in the past (van Atteveldt & Peng, 2018). Rich semantic information in digital traces can provide a social telescope (Golder & Macy, 2014) to observe or infer *what* is produced, shared, and consumed by ordinary users. Multiple social and interactive relations in digital traces facilitate empirical studies on *who* connects with *whom* in various contexts, such as political campaigns, information diffusion, and health behavior changes. Timestamps, another genetically built-in element in digital traces, include significant detailed information on *when* human behaviors occur. Due to the tight interconnection between social/mobile media and everyday life, digital traces on social/mobile media can not only be used to infer information about events occurring on social/mobile media themselves, but also about behaviors performed by its users in the physical world.

Human behaviors, either online or offline, take place in time and use up time. "If the unity of all science is an elusive (and perhaps illusory) goal, it is nevertheless true that there are some concepts that, either in their mathematical or conceptual formulations, are common to a number of scientific fields" (McDonald & Dimmick, 2003, p. 60). Time is one such concept, which has attracted scholarly attention in various disciplines, such as sociology (Sorokin & Merton, 1937), political science (Hall, 2016; Cohen, 2018), management (Ancona et al., 2001a), and communication (Zhu et al., 2018; Peng & Zhu, 2020; Peng et al., 2020). The rich temporal information in digital traces and the advancement of temporal modeling algorithms provide a microscope with renovated temporal lens for social scientists to observe and examine human behaviors. The current paper will discuss what are the properties of timestamps in digital traces, how timestamps in digital traces have been utilized in existing studies of online behaviors, and what are potential perspectives to explicate the meaning of time in online behaviors, and what are future directions to make sound use of timestamps in computational social science.

## 2   (Un)Desirable Properties of Timestamps in Digital Traces

Timestamps in digital traces possess certain desirable properties for empirical examination of online behaviors. First, the timestamp is *universally available* in almost all types of digital traces. Time is the (only) common denominator of all types of digital traces, ranging from webpage visit logs, to email communication, online forum posts, Twitter posting/retweeting, video sharing/commenting, following behavior, and calling/texting behavior and app use behavior on mobile phones. Timestamps in digital traces have *fine-grained granularity*. Digital traces offer a much finer timescale (daily, hourly, or even per second), which makes it possible to study the subtle nuance underlying online behaviors. Moreover, timestamps in digital traces are *universally standardized*, which enables the comparison of time and time-related constructs between individuals, between platforms, and between past and present.

Timestamps in digital traces are *reliable* and *accurate*. Owing to the unobtrusive nature of digital traces (Salganik, 2019), temporal information captured in digital traces has been found to possess better quality than self-reported measures (e.g., Boase & Ling, 2013; Araujo et al., 2017), as they can substantially correct social desirability bias, memory bias, and other types of bias found in self-reported measures (van Atteveldt & Peng, 2018; Salganik, 2019). Timestamps in digital traces is *naturally interpretable* and *conceptually straightforward*, which makes theoretical constructs derived from timestamps in digital traces possess very *strong face validity*.

Like other types of data, timestamps in digital traces also have some undesirable properties which constrains its potential in empirical studies of online behaviors. Although the timestamp is universally available, most of digital traces available to social scientists have *narrow coverage* which include online behaviors from a single platform or an individual application. Due to ethical and privacy concerns, it is very challenging, if not completely impossible, to link online behaviors across platforms at individual level. Secondly, despite the streaming nature of digital traces, most of the digital traces available to social scientists have very *insufficient duration*, which can cover online behaviors under study for several weeks or months. This poses great challenges for social scientists to uncover the full dynamics underlying online behaviors from adoption to abandonment.

# 3 How are Timestamps in Digital Traces Utilized in Empirical Analysis of Online Behaviors?

As a very fundamental dimension underlying all human behaviors, time is considered as a primitive term (Chaffee, 1991) in behavioral studies across all disciplines of social science. It has been incorporated into empirical studies of human behaviors and beyond. The foremost consideration of time is to employ time as a measurement instrument of human behaviors and related constructs. Fine-grained temporal information in digital traces have been employed to measure different behaviors in empirical studies, such as general media use (Taneja et al., 2012), news exposure (Zhang et al., 2017), and health tracking behavior on mobile devices (Guan et al., 2019).

Secondly, directly observed timestamp in digital traces has been employed to infer users' unobservable characteristics in digital traces. Despite the large volume of digital traces in computational social science, digital traces contain very limited information about users' characteristics, which are very important antecedents or outcome variables in social science. How individuals allocate time to different activities (e.g., leisure activities and work activities) is found to be related to their quality of life (Greenhaus et al., 2003), social status and lifestyle (Bellezza et al., 2017). Thus, timestamps of digital traces have been employed to infer personal characteristics of thousands or millions of users, such as employment status (Bokányi et al., 2017) and occupation (e.g., Jiang et al., 2012). This can substantially improve the values of digital traces in computational social science.

Thirdly, timestamps in digital traces have been employed as a medium to study change of online behaviors (Zhang et al., 2017) or as a causal link between variables (Sun et al., 2014) in computational social science. How mobile news applications will influence the diversity of news consumption has attracted much scholarly attention. With a large-scale behavioral dataset extracted from a mobile news application, Zhang et al. (2017) found that the diversity of consumed news on mobile devices declines over time. Such declining trend in the diversity of mobile news consumption is more remarkable among female users than among male users. It has been a long-standing concern if and how social issues on public agenda will interact with one another to gain public attention. By analyzing about 450 million tweets on 10 social issues posted by American users in 12 months, Sun et al. (2014) revealed that social issues can compete and cooperate with one another to gain public attention. The cooperation between social issues is driven by the similarity between social issues.

# 4 Explicate the meaning of time in online behaviors: Behaviors-in-Time versus Time-in-Behaviors

Social scientists in all subject areas are using the concept of time explicitly or implicitly in research design, measurements, and/or data analysis, but very few are talking about it. The meaning of time is under-explicated in empirical studies of online behaviors. It is a common sense that time is a scarce resource that can be used and allocated by the public (Adam, 1990). However, time is not just an objective quantity on clocks and calendars. It is a fact of life which represent ordering principle and force for selection and prioritizing and can be used as a tool for coordination, orientation, and control (Elias, 1984). To better explicate the meaning of time in online behaviors, it is essential and desirable to elucidate how time and human behaviors can be related to one another.

## *4.1 Behaviors-in-Time Perspective*

Behaviors-in-time is a dominant perspective on the relationship between time and online behaviors. It argues that individuals are consumers of time. Time "exists independently of human actions, and is thus experienced as a powerful constraint on those actions" (Orlikowski & Yates, 2002, p. 68). Time provides a boundary within which human behavior is enacted (Clark, 1985). Time shapes human behaviors. Under this behaviors-in-time perspective, time is homogeneous, uniform, regular, precise, and deterministic (Ancona et al., 2001a).

The behaviors-in-time perspective is characterized by their quest for precision, with time as a resource being expressed by numerical specification of both the duration and the frequency with which behaviors are carried out. The behaviors-in-time perspective generates a proliferation of research and largely continues along a one-dimensional track. It has resulted in the time-budget paradigm in empirical studies and led to a prevalence measurement of online behaviors. The prevalence measurement projects individuals' behaviors (e.g., texting, communication, news consumption, social networking, watching videos, listening to music) to a time continuum and reduce human behaviors to a prevalence measurement on the continuum (e.g., Boase & Ling, 2013; Gerpott & Thomas, 2014; Vanden Abeele et al., 2013), such as the number of unique behaviors on the continuum, the repeated occurrences of such behaviors, the duration of a particular behavior or all behaviors as a whole, or the (in)equality in time allocation across different behaviors. The behaviors-in-time perspective is conceptually intuitive; thus it has been widely applied in empirical quantification of online behaviors. Despite its popularity, the behaviors-in-time perspective suffers from two problems, which pose great hurdles to our understanding of intricate patterns underlying online behaviors.

First, the meaning of time is over-simplified, if not completely forgotten, in the behaviors-in-time perspective. The behaviors-in-time perspective and the time-budget paradigm focus on how individuals allocate their time to different behaviors in daily lives. It "excludes all other aspects of time that might simultaneously have a bearing on people's lives, and that people relate to, at any one moment" (Adam, 1990, p. 95). However, time should be a multi-dimensional phenomenon (Flaherty, 2003) implying far more than the prevalence of various behaviors executed by individuals. Secondly, the over-simplified meaning of time in the behaviors-in-time perspective leads to an under-estimation of users' locus of control underlying various types of human behaviors. Individuals not only *consumers* of their time but also *owners* of their time (Peng & Zhu, 2020). As time reflects, regulates, and orders social life, time is always considered a social fact which is actively constructed by individuals in a society (Sorokin & Merton, 1937). The behaviors-in-time perspective fails to adequately capture individuals' activeness or locus of control (Lefcourt, 1991) in online behaviors. With the locus of control, individual users are entitled to bend social and mobile media to their own use (Biocca, 1988). The locus of control within individual users will not only drive them to budget their time for various behaviors, but also empower them to actively create individualized repertoire of behaviors and array selected behaviors into different sequential processes to gratify their different needs.

Time is not only a fundamental organizing principle underlying online behaviors, but a plural with multiple possibilities (Brunelle, 2017). With the rapid advancement of information and communication technologies, users are entitled to engage in a proliferating pool of behaviors on social and mobile media, implying users have greater freedom to determine how and when they execute different types of behaviors and to develop personalized behavioral patterns on their own. Timestamps genetically built into digital traces on social and mobile media make it empirically feasible to restore those behavioral patterns. It calls for a microscope with a renovated and sharpened temporal lens in empirical studies of online behaviors.

## 4.2 *Time-in-Behaviors Perspective*

Time-in-behaviors perspective is such a renovated temporal lens that enables an intricate re-explication of the meaning of time in online behaviors. The time-in-behaviors perspective argues that individuals are owners of time which comes from the behaviors that are correlated with each other (Sorokin & Merton, 1937). Human behaviors are planned in relative to other behaviors (Lauer, 1981), and individuals transition from one to the next when they internally sense that the former behavior is complete and the next behavior is relevant to and compatible with the former behavior (Avnet & Sellier, 2011). Time is shaped in human behaviors, which is considered as relative, contextual, organic, and socially constructed (Adam, 1990; Orlikowski & Yates, 2002).

The time-in-behaviors perspective adequately recognizes the agency of human being in scheduling online behaviors. The selection, order, and transition in online behaviors, which have been completely ignored in the behaviors-in-time perspective, can lead to the emergence of a three-layer hierarchical structure of human behaviors, as illustrated in Figure 1.

Figure 1 about here

The micro-level layer in this hierarchical structure is a wide variety of discrete behaviors with which an individual user can get engaged. Driven by diverse personal needs and constrained by limited time resources, users will develop a self-defined repertoire of behaviors (Taneja et al., 2012) out of all available behaviors in online platforms, ranging from computer-mediated communication to web surfing, social networking, entertainment, playing games, and many others. The behaviors-in-time perspective focuses on how individuals will allocate their time resources to discrete behaviors in this micro-level layer, which has resulted in many theories in social science, such as uses and gratification theory (Blumler et al., 1985) and media richness theory (Daft & Lengel, 1986).

Each behavior included in a self-defined repertoire is not completely independent and distinct from one another. Instead, to satisfy individuals' different needs in daily lives, individual behaviors can be organized and ordered into a consecutive uninterrupted sequence of behaviors, which is conceptualized as behavioral sessions in empirical studies (van Canneyt et al., 2017; Zhu et al., 2018; Peng & Zhu, 2020). Behavioral sessions constitute the meso-level layer in the hierarchical structure. Some sessions may include a variety of consecutive behaviors among which a user can switch back and forth, while others may include one singular type of behavior which a user will stick to in a certain timespan. Some of the behaviors in a session can be enduring, while others in the same session may be quite ephemeral.

Behavioral sessions are not single and isolated occurrences in online platforms. Behavioral sessions are inter-dependent on one another, which requires researchers to view behavioral sessions as paced, meaningful, higher-order sequences of events (Ancona et al., 2001a). Behavioral sessions can be concatenated into the macro-level layer in the hierarchical structure – behavioral trajectories. Some behavioral trajectories are quite varied or complex: Users can develop many distinct behavioral sessions and experience transitions among those behavioral sessions. Others, within the same time window, will stay in one behavioral session for the full observation period, and therefore, such trajectories are argued to be relatively "simple" or "stable".

## *4.3 Quantifying Behavioral Sessions: Challenges and Opportunities*

Behavioral sessions act as a meso-level behavioral pattern which is composed of micro-level behaviors and is also a component of macro-level behavioral trajectories. Behavioral sessions constructed under the time-in-behaviors perspective explicitly and deliberatively creates a sequential order among those behaviors, which means a richer set of temporal dimensions, such as timing, pace, rhythms, and transition (Ancona et al., 2001b), to characterize individuals' online behaviors. It well preserves the process nature of online behaviors and considers the full spectrum of information incorporated in behavioral sessions, such as total and consecutive time spent on each behavior and the order and transition between different behaviors. Here we review how behavioral sessions can be constructed from timestamped digital traces and what research questions can be answered with behavioral sessions.

### 4.3.1 How to construct behavioral sessions from timestamped digital traces?

A significant challenge to study behavioral sessions with digital traces is how to construct a behavioral session from digital traces. Although digital traces include detailed timestamp about each individual behaviors in online platforms, there lacks obtrusive stamps about the start and end of behavioral sessions. In other words, behavioral sessions need to be constructed from digital traces.

There are two general approaches to construct behavioral sessions from digital traces: the threshold approach and screen-based approach. The threshold approach defines a behavioral session based on a threshold determined by the duration of inactiveness, which has been used to study behavioral sessions in webpage browsing (e.g., Arlitt, 2000; Mehrzadi & Feitelson, 2012) and mobile app use (Böhmer et al., 2011; van Canneyt et al., 2017). The screen-based approach is a newly developed approach suitable to construct behavioral sessions on mobile devices (e.g., Falaki et al., 2010; Zhu et al., 2018). The screen-based approach will define a session based on the (de)activation of the screens of mobile devices, by assuming that such (de)activation of screens is a valid indicator of intentional human behavior on mobile devices.

Based on the development of empirical research on human dynamics (Barabasi, 2005), we proposed an individualized threshold approach to construct behavioral sessions (Peng & Zhu, 2020). It has been found in human dynamics research that there is a power-law distribution of inter-behavior time between consecutive behaviors in different domains (Peng et al., 2020). It implies that individuals will schedule different behaviors on mobile phones based on the priority assigned to different tasks (Barabasi, 2005) or their perception of past activity rate (Vazquez, 2007). When individuals perceive two discrete behaviors (e.g., search and email) to have high priority or relate with one another, individuals will execute the two behaviors with very short inter-app interval, implying that the two discrete behaviors can be assigned into one behavioral session. Moreover, different individuals may have different priority-assignment mechanism or different perception of their activity rates. Therefore, the median value of inter-behavior intervals is employed as an individualized threshold to construct behavioral sessions for each user from the records of discrete app use behaviors on mobile devices. The inter-behavior interval approach proposed in Peng & Zhu (2020) is grounded in well-articulated theoretical underpinnings, which does not require any exogenous information to construct behavioral sessions. It has the potential to be applied to any timestamped digital traces on social and mobile media.

### 4.3.2 What can be done with behavioral sessions?

The strength of time-in-behaviors perspective lies in that it allows us to go beyond a single behavior and construct higher-order sequence of behavior to quantify human behaviors. We can transcend the taken-for-grantedness in existing quantification of online behaviors and explore some new perspectives in explicating online behaviors. Figure 2 summarizes some potential dimensions that can be explored to explicate online behaviors underlying both behaviors-in-time and time-in-behaviors perspectives.

---

Figure 2 about here

---

The order and transition in behavioral sessions reintroduces chronology into the studies of human behaviors and empowers researchers to capture the multi-dimensional meaning of time in empirical observations of online behaviors. Moreover, behavioral sessions acknowledge the embeddedness of the past and the future in the present, implying that "it may not be meaningful to dissect time and artificially separate it in to discrete, objective units to which we assign specific content" (George & Jones, 2000, p. 660).

In our study of mobile applications ("mobile apps" hereinafter) use (Peng & Zhu, 2020), we employed 11.5 million records of discrete mobile apps use by a representative panel of approximately 2,500 users in Hong Kong. With the aforementioned median inter-behavior interval approach, we constructed 3.7 million behavioral sessions for mobile app use. Behavioral sessions enable the examination of the transition between different types of mobile apps. It is found that the assortative transition, referring to a transition which originates from and terminates on the same category of mobile apps, substantially outweighs the disassortative transition referring to a transition that originates from one category of app and terminates on another category app (Peng & Zhu, 2020). The average disassortative transition rates between the 21 categories of mobile apps are illustrated in Figure 3. The transitions between different types of mobile apps implies that individuals can act as agent to combine different types of online behaviors into new behavioral patterns to gratify their needs.

Theoretically, individuals can develop a large amount of possible sequential combination of online behaviors. Are there any regularities underlying the sequential combination of online behaviors? The advancement of sequence alignment algorithms and the development of sequence-level metrics make it possible to employ a whole-sequence approach, rather than a point-to-point transition approach, to study online behaviors. The whole-sequence perspective is adopted to examine whether different behavioral sessions are the same or not, how similar they are, or whether parts are similar. An individualized approach is adopted in Peng & Zhu (2020) to align mobile sessions and detect sequential patterns underlying mobile sessions considering the possible individual differences on sequential patterns underlying mobile sessions. In total, 2,203 behavioral sessions, which represent sequential patterns underlying 1.8 million multi-app behavioral sessions, are identified for all users in the panel. The reduction of 1.8 million behavioral sessions to 2,203 representative sequential patterns implies that there exist great regularities in behavioral sessions underlying mobile phone use.

---

Figure 3 & Figure 4 about here

---

Behavioral trajectories, which is the macro-level organization of online behaviors in the hierarchical structure of online behaviors, can vary in terms of the number of distinct behavioral sessions, the order of behavioral sessions, and the variance of durations spent on different behavioral sessions (Aisenbrey & Fasang, 2010). The variation embedded in a behavioral trajectory can be quantified as sequential complexity (Elzinga, 2010) which provides a summary measure of the variability within individual behavioral trajectory. Based on the data in Peng & Zhu (2020), we found that there is a zigzag pattern on the circadian rhythm of sequential complexity underlying behavioral trajectories, as shown in Figure 4. Users will kick off a day with a behavioral trajectory which has relatively low complexity score during small hours. The complexity will increase to a significantly higher degree during morning hours, which will then dive to the lowest level of a day during mid-day hours. After that, users will organize their behavioral sessions in a much more complex way during afternoon hours, which will leap to the peak of a day during evening hours. Moreover, users are found to have a more complex organizations of mobile sessions on workweeks than on weekends.

Behavioral sessions consider human behaviors as an unfolding process over time. It goes beyond duration/frequency and offers a new scale to quantify online behavior. Duration is the most used time metric to quantify online behaviors. The duration metric aggregates the time an individual spends on a behavior per given unit (e.g., a day). It is an equal-length scale which consider each minute being the same length (Zhu et al., 2018). Despite the obvious advantage of such equal-length metrics in quantifying online behaviors (e.g., high face validity, naturally interpretable, and conceptually simple), such equal-length scale fails to capture all temporal characteristics of online behaviors. An alternative scale called "session" is proposed to quantify online behaviors. Session is a continuous uninterrupted sequence of behaviors an individual performs on a mobile device (Zhu et al., 2018). Each session can last differently, so it is considered as a variable-length scale. Different from the equal-length time metrics, a session can capture the full range of the temporal characteristics of online behaviors, including duration, frequency, timing, and sequence. Zhu et al. (2018) empirically demonstrated that session provide additional information about online behaviors over and above the equal-length metrics.

The proposed variable-length scale enables empirical tests of long-standing theoretical concerns in social science. For example, will the increasing integration of mobile device into daily lives drives time use more fragmented? The equal-length scale could not adequately answer this question, as fragmentation is not about the duration of online behaviors, but about the pattern (e.g., frequency and sequence) of their online behaviors (Zhu et al., 2018). The variable-length scale incorporates sufficient information to address this concern.

Meanwhile, as timing is incorporated into variable-length metrics, it provides a significant opportunity to study the repetitiveness or habit underlying online behaviors. Time is cyclically repetitive, which is often unaccounted for or under-utilized in existing studies of online behaviors. Identifying their frequency, rhythm, and cycles is often a key to understanding the nature of online behaviors (George & Jones, 2000). Despite the widely accepted notion of habit or repetitiveness in human behaviors (Wood et al., 2002), how to quantify the repetitiveness or habit underlying human behaviors has remained a long-standing challenge in social science. Repetitive behaviors on mobile devices can serve as a valid signal of temporal regularities. Zhu et al. (2018) defined temporal regularities of online behaviors as repetitive mobile sessions occurring in a 24-hour cycle, which is measured by an innovative indicator called Rate of Repeated Sessions (RRS). RRS describes the degree of temporal regularity for a user to engage in a mobile session at fixed time slots. It is found that nearly half of online behaviors on mobile devices happen at fixed time slots, implying that individuals' online behaviors on mobile device "appear to be *prescheduled* (i.e., predictably recurring) rather than impulsive or random" (Zhu et al., 2018, p. 226).

# 5   It's Time to Talk about Time in (Computational) Social Science

Time is an essential dimension underlying human behaviors, whether online or offline. It makes little sense to ignore it, treat it implicitly, or treat it in an inadequate manner (George & Jones, 2000). Several recently published meta-analysis (e.g., Liu et al., 2019; Hancock et al., 2019) has found that duration or frequency of online behaviors has zero or limited effects on psychological well-being. In other words, behaviors-in-time perspective may increase the chance of Type II error in empirical examination of the effects of online behaviors. In computational social science, time and temporality underlying online behaviors should and can move from the background to the foreground. Researchers need to move beyond the behaviors-in-time perspective, capture everything they do, and restore the complexity behind the selection and organization of different behaviors in the temporal dimension (Reeves et al., 2020). Collective efforts from government, academia, and commercial companies are needed to harness multiple sources of digital traces across platforms, which can enable a panoramic examination of online behaviors in exquisite detail. Although there are a lot of work to be done in this regard, the following directions are worth our immediate attention in computational social science.

## 5.1   *Different behavioral processes across different granularities of time unit*

The most frequent and most intuitive use of time in computational social science is to use it as a "grouping" variable by which the unit (i.e., duration) of analysis for given behavioral variable(s) is/are defined. For example, to quantify the degree of engagement of a user on social media (e.g., Twitter), we usually count the number of tweets the user posted or forwarded during a certain time unit, e.g., per day, week, month, etc. While the overall/long-term trend in the given data series is likely to remain similar across different time units, different patterns may emerge in short-/mid-terms. Unfortunately, we know very little, if any, about correlative patterns, let alone causal mechanisms, between time units and human behaviors at this point, but certainly should fill the gap in this line of research that is both theoretically informative and methodologically useful.

## 5.2  Multiple ways to use time as an explanatory variable

Computational social science is largely interested in explanatory research (Hofman et al., 2021). However, most of the data available to computational social scientists are readymade records (e.g., user generated content or user browsing logs). Time has often been used as a grouping variable to define the unit of analysis (see above) or a covariate to control for trend effects in time series analysis. In fact, time can offer more for explanatory purpose, by serving an explicitly operationalized independent variable. One common approach in traditional social science is to create dummy variable(s) for interruption/intervention event(s) based on the relevant time point(s) and then examine its/their causal impact on subsequent behavior(s). Another approach, though less frequently used, is to construct "temporal profile" for each user (in individual-level analysis) or each event (in aggregate-level analysis) during the pre-event period and then examine its causal impact on relevant behavior(s) in the post-event period. There could be many other ways to operationalize time as an independent variable, depending on specific theorical grounds or research purposes. A common strategy underlying these approaches is to come out of the box of classic time series analysis, which appears to be more powerful for descriptive/predictive research than explanatory research.

# 6  Further Readings

Peng, T. Q., & Zhu, J. J. H. (2020). Mobile phone use as sequential processes: From discrete behaviors to sessions of behaviors and trajectories of sessions. *Journal of Computer-Mediated Communication, 25*(2), 129-146. doi:10.1093/jcmc/zmz029

Zhu, J. J. H., Chen, H., Peng, T. Q., Liu, X. F., & Dai, H. (2018). How to measure sessions of mobile phone use? Quantification, evaluation, and applications. *Mobile Media & Communication, 6*(2), 215-232. doi:10.1177/2050157917748351

Figure 1 Hierarchical Structure of Online Behaviors

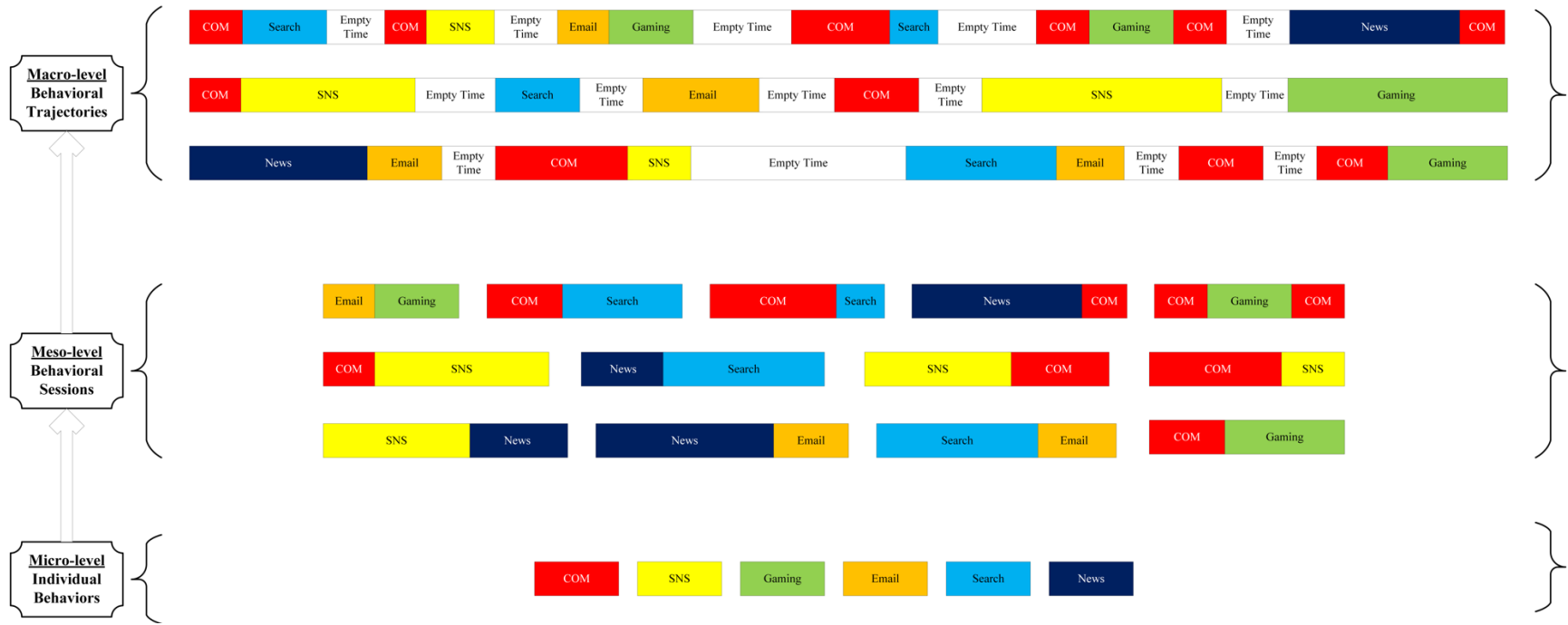

Note: COM = Online Communication; SNS = Online Social Networking Behavior; Gaming = Online Gaming Behavior; Email = Email Communication; Search = Online Information Search; News = Online News Consumption.

Figure 2 Multi-dimensional Explication of Time in Online Behaviors

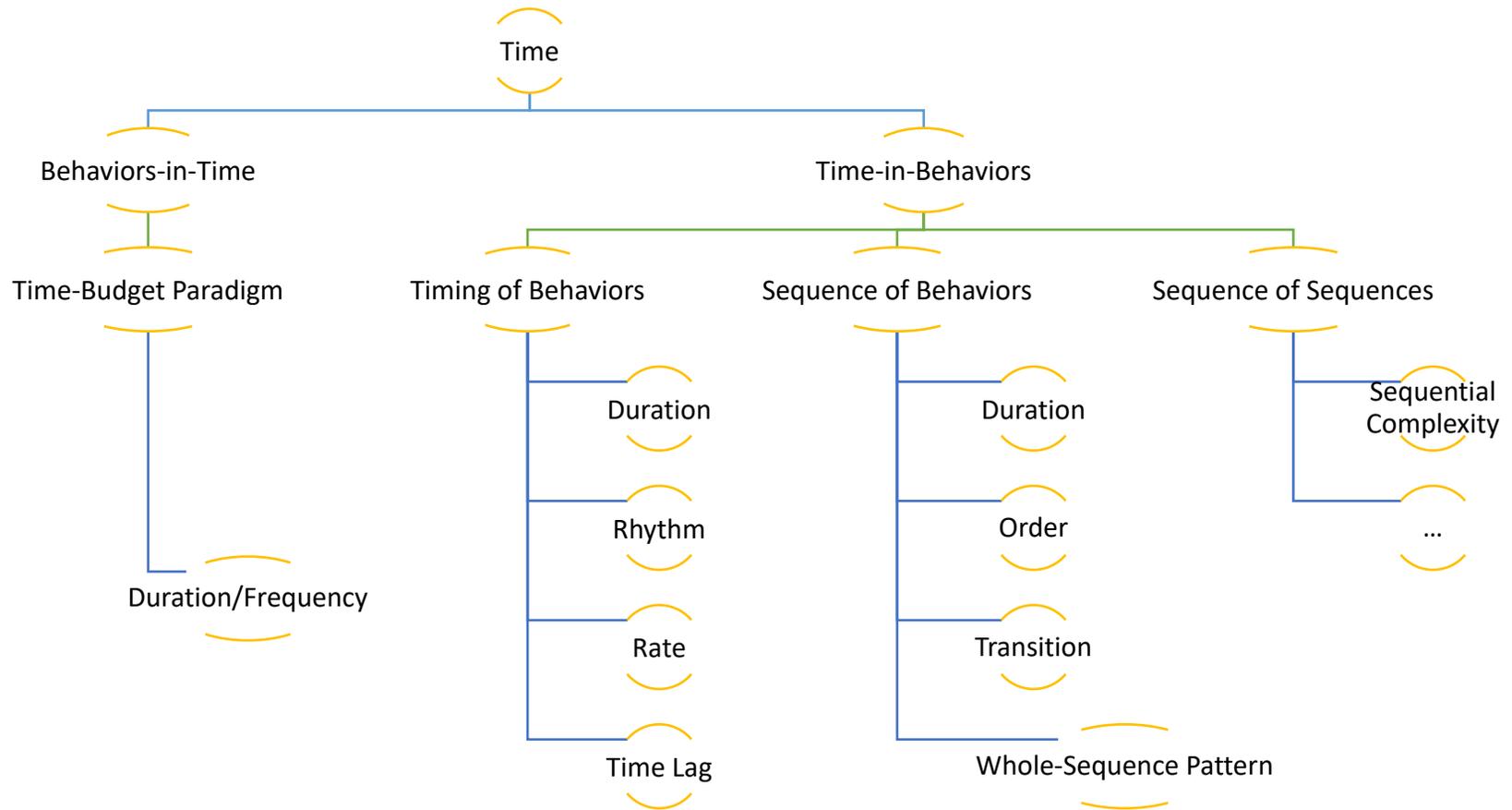

Figure 3 Transition between Mobile Applications in Behavioral Sessions

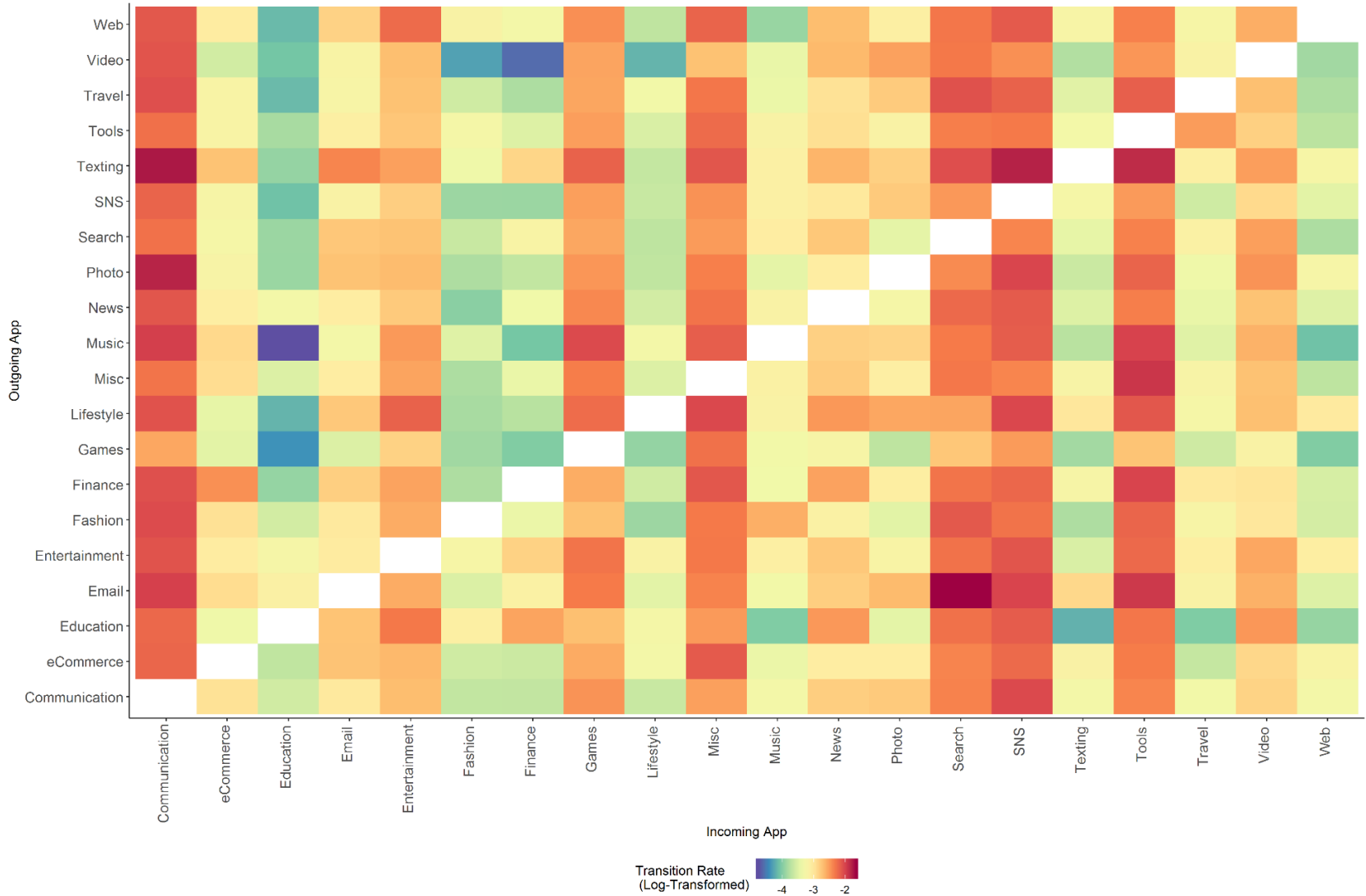

Data Source: Peng & Zhu (2020)

Figure 4 Circadian Rhythm of Sequential Complexity of Behavioral Trajectories on Mobile Devices Adjusted by Weekdays

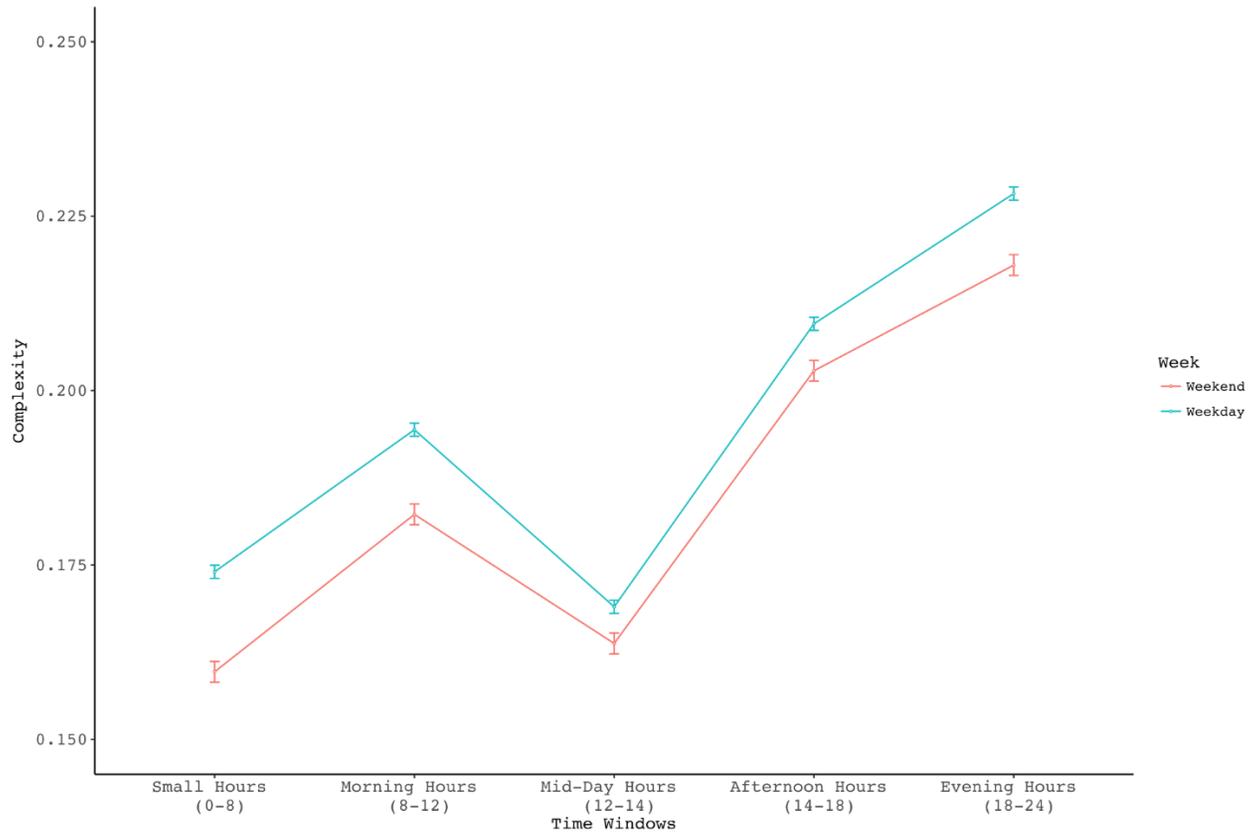

Data Source: Peng & Zhu (2020)